\UseRawInputEncoding

\documentclass[reprint,prd,amsmath,amssymb,aps,floatfix,twocolumn,nofootinbib]{revtex4-1}

\usepackage[english]{babel}
\usepackage{graphicx}
\usepackage{amssymb,amsmath}
\usepackage{xcolor}
\usepackage{hyperref}
\usepackage{aas_macros}

\begin{document}


\title{Evidence for the transition of a Jacobi ellipsoid into a Maclaurin spheroid in gamma-ray bursts}

\author{J. A. Rueda$^{1,2,3,4,5}$, R. Ruffini$^{1,2,6}$, L. Li$^{1,2,7}$, R. Moradi$^{1,2,7}$, J. F. Rodriguez$^{8,2}$, Y. Wang$^{1,2,7}$}

\affiliation{$^{1}$ICRA, Dipartimento di Fisica, Sapienza Universit\`a  di Roma, Piazzale Aldo Moro 5, I-00185 Roma, Italy}
\affiliation{$^{2}$ICRANet, Piazza della Repubblica 10, I-65122 Pescara, Italy}
\affiliation{$^{3}$ICRANet-Ferrara, Dipartimento di Fisica e Scienze della Terra, Universit\`a degli Studi di Ferrara, Via Saragat 1, I--44122 Ferrara, Italy}
\affiliation{$^{4}$Dipartimento di Fisica e Scienze della Terra, Universit\`a degli Studi di Ferrara, Via Saragat 1, I--44122 Ferrara, Italy}
\affiliation{$^{5}$INAF, Istituto di Astrofisica e Planetologia Spaziali, Via Fosso del Cavaliere 100, 00133 Rome, Italy}
\affiliation{$^{6}$INAF, Viale del Parco Mellini 84, 00136 Rome, Italy}
\affiliation{$^{7}$INAF, Osservatorio Astronomico d'Abruzzo,Via M. Maggini snc, I-64100, Teramo, Italy}
\affiliation{$^{8}$
Escuela de F\'isica, Universidad Industrial de Santander, Ciudad Universitaria, Bucaramanga 680002, Colombia}

\email{jorge.rueda@icra.it; ruffini@icra.it; rahim.moradi@inaf.it;...}

\date{\today / Received date /
Accepted date }

\begin{abstract}
In the binary-driven hypernova (BdHN) scenario, long gamma-ray bursts (GRBs) originate in a cataclysmic event that occurs in a binary system composed of a carbon-oxygen (CO) star and a neutron star (NS) companion in close orbit. The collapse of the CO star generates at its center a newborn NS ($\nu$NS), and a supernova (SN) explosion. Matter from the ejecta is accreted both onto the $\nu$NS because of fallback and onto the NS companion, leading to the collapse of the latter into a black hole (BH). Each of the ingredients of the above system leads to observable emission episodes in a GRB. In particular, the $\nu$NS is expected to show up (hereafter $\nu$NS-rise) in the early GRB emission, nearly contemporary or superimposed to the ultrarelativistic prompt emission (UPE) phase, but with a different spectral signature. Following the $\nu$NS-rise, the $\nu$NS powers the afterglow emission by injecting energy into the expanding ejecta leading to synchrotron radiation. We here show that the $\nu$NS-rise and the subsequent afterglow emission in both systems, GRB 180720B and GRB 190114C, are powered by the release of rotational energy of a Maclaurin spheroid, starting from the bifurcation point to the Jacobi ellipsoid sequence. This implies that the $\nu$NS evolves from a triaxial Jacobi configuration, prior to the $\nu$NS-rise, into the axially symmetric Maclaurin configuration observed in the GRB. The triaxial $\nu$NS configuration is short-lived (less than a second) due to a copious emission of gravitational waves, before the GRB emission, and it could be in principle detected for sources located at distances closer than $100$ Mpc. This appears to be {a specific} process of emission of gravitational waves in {the BdHN I powering long GRBs}.
\end{abstract}
\maketitle


\section{Introduction} \label{sec:intro}

The progenitor of long gamma-ray bursts (GRBs) in the binary-driven hypernova (BdHN) model is a binary system composed of a carbon-oxygen (CO) star and a neutron star (NS) companion (see, e.g., Refs. \cite{2012ApJ...758L...7R, 2012A&A...548L...5I, 2014ApJ...793L..36F, 2015PhRvL.115w1102F, 2015ApJ...812..100B, 2016ApJ...833..107B, 2019ApJ...871...14B}). The gravitational collapse of the iron core of the CO star leads to the formation of a newborn NS ($\nu$NS) at its center and to a supernova (SN) explosion that eject the outer layers of the star. Some part of the ejecta is accreted by the NS companion and also by the $\nu$NS via matter fallback accretion. Both accretion processes proceeds at hypercritical (i.e., highly super-Eddington) rates thanks to a copious neutrino emission \citep{2016ApJ...833..107B, 2018ApJ...852..120B}. In compact binaries with orbital periods of a few minutes, the hypercritical accretion onto the NS companion brings it the critical mass inducing its gravitational collapse and forming a rotating (Kerr) BH. These systems have been called BdHN I. In less compact binaries, the NS companion does not reach the critical mass and hold stable as a more massive, fast rotating NS. These systems have been called BdHN II. 

In this article, we focus on BdHN I. The fallback accretion onto the $\nu$NS spins it up to a millisecond rotation period {(see \cite{2019ApJ...871...14B} for numerical simulations). We shall show in this work that in} this early phase, the huge $\nu$NS rotational energy of up to a few $10^{53}$ erg can power what we have called the $\nu$NS-rise, i.e., the first observed emission from the $\nu$NS. Subsequently, the $\nu$NS fuels the synchrotron radiation {originated from the expanding SN ejecta leading} to the afterglow observed in the X-rays, optical and radio energy bands following a power-law luminosity \cite{2018ApJ...869..101R, 2020ApJ...893..148R}. The accretion onto the companion NS leads to the BH formation that leads to the ultrarelativistic prompt emission (UPE) phase \cite{2021PhRvD.104f3043M} and the GeV emission \citep{2019ApJ...886...82R, 2020EPJC...80..300R, 2021MNRAS.504.5301R, 2021A&A...649A..75M, 2022ApJ...929...56R}. 

{We aim here to estimate the $\nu$NS energy budget and check if it explains the $\nu$NS-rise emission. We use as a proxy the case of GRB 180720B and GRB 190114C. For this task, we first individuate in the data of these sources the $\nu$NS-rise emission based on the expectation of the BdHN I model. First, if the $\nu$NS powers the $\nu$NS-rise and the X-ray afterglow, we look for the conjunction of the back-in-time extrapolation of the observed X-ray power-law luminosity of the afterglow with the $\nu$NS-rise power. Second, we expect the $\nu$NS-rise to show up either as a precursor to the UPE or at most to superpose to it. Having establish the connection between the $\nu$NS-rise and the afterglow, and with the knowledge of their energetics, we evaluate if the $\nu$NS can indeed power these emissions. For self-consistency with the modeling of the afterglow in the BdHN scenario (see, e.g., \cite{2018ApJ...869..101R, 2020ApJ...893..148R}), we seek for the $\nu$NS parameters that can explain the emissions demanding rigid rotation and axial symmetry. With the estimate of the $\nu$NS parameters, we discuss the previous early life of the $\nu$NS and on the possible associated emission of gravitational waves during its evolution towards the axially-symmetric stage.}

{We describe in section \ref{sec:II} the sequence of physical phenomena that occur in a BdHN I and their associated observables in the GRB data. In section \ref{sec:III}, we first individuate the $\nu$NS-rise emission in GRB 180720B and GRB 190114C. Then, we proceed to the evaluation of the $\nu$NS parameters and their evolution modeling it as a stable Maclaurin spheroid. This assumption, together with the energy conservation equation, allows to infer the time evolution of the $\nu$NS structure without additional assumptions. We show} that the $\nu$NS-rise and the afterglow emission energetics require that the initial parameters of the Maclaurin spheroid {are close to the parameters} of the bifurcation point to the Jacobi sequence of ellipsoids. {This result suggests that} the $\nu$NS before the GRB emission evolves from a Jacobi ellipsoid into a Maclaurin spheroid by emission of gravitational waves. {Therefore, the only mechanism that can generate gravitational radiation in the BdHN scenario of GRBs originates in the transition from the triaxial configuration (ellipsoid) at birth to the axially-symmetric state (spheroid). In section \ref{sec:IV}, we elaborate on the entity of this emission and discuss its possible detectability.}

{\section{Sequence of BdHN physical phenomena and observables}} \label{sec:II}

{The above sequence of physical phenomena that occur in a BdHN I are related to specific observational episodes in the GRB data that we summarize in Table \ref{tab:observables} and discussed below for GRB 180720B and GRB 190114C.}

\begin{table*}
    \centering
    {
    \caption{GRB observables associated with the BdHN I component and physical phenomena. References in the table: $^a$\cite{2019ApJ...874...39W},$^b$\cite{2014ApJ...793L..36F, 2016ApJ...833..107B}, $^c$\cite{2019ApJ...886...82R, 2021A&A...649A..75M, 2021PhRvD.104f3043M},     $^d$\cite{2001A&A...368..377B, 2021PhRvD.104f3043M}, $^e$\cite{2019ApJ...886...82R, 2020EPJC...80..300R, 2021A&A...649A..75M, 2022ApJ...929...56R}, $^f$\cite{2018ApJ...852...53R}, $^g$\cite{2018ApJ...869..101R, 2019ApJ...874...39W, 2020ApJ...893..148R}}
    \begin{tabular}{l|c|c|c|c|c}
    \hline
     BdHN I component/phenomena &  \multicolumn{5}{c}{GRB observable} \\
     \cline{1-6}
          & $\nu$NS-rise & UPE & GeV & X-ray flares & Afterglow\\
          &  (soft-hard X-rays)       & (MeV) & emission & early afterglow & (X/optical/radio)\\
          \cline{1-6}
    Early SN emission$^a$ & $\bigotimes$ & & & & \\
    \cline{1-1}
    Hypercritical accretion onto the $\nu$NS and NS$^b$  &  $\bigotimes$ & & & & \\
      \cline{1-1}
     BH formation from NS gravitational collapse$^c$ & & & $\bigotimes$ & & \\ 
     \cline{1-1}
    Transparency of ultrarelativistic $e^+e^-$ (from vacuum & & $\bigotimes$ & & &\\
    polarization) in low baryon load region$^d$  & & & & & \\
     \cline{1-1}
    Synchrotron emission by the \textit{inner engine}:  & & & $\bigotimes$ & &\\
    newborn BH + $B$-field+SN ejecta$^e$ & & & & & \\
     \cline{1-1}
    Transparency of ultrarelativistic $e^+e^-$ (from vacuum & & & & $\bigotimes$ &\\
    polarization) in low baryon load region (SN ejecta)$^f$  & & & & &\\
     \cline{1-1}
    Synchrotron emission from SN ejecta with & & & & & $\bigotimes$\\
    energy injection from $\nu$NS$^g$ & & & & & \\
     \cline{1-1}
    Pulsar-like emission from the $\nu$NS$^g$ & & & & & $\bigotimes$\\
    \cline{1-6}
    \end{tabular}
    }
    \label{tab:observables}
\end{table*}

\textit{\textbf{The UPE phase}}. The BH forms and together with the surrounding magnetic field and ionized matter from the SN ejecta composes the \textit{inner engine} \cite{2019ApJ...886...82R}. The gravitomagnetic interaction of the newborn Kerr BH with the magnetic field induces an electric field \cite{2020EPJC...80..300R, 2022ApJ...929...56R}. The electric field is initially overcritical, i.e., larger than the quantum electrodynamics (QED) critical field for vacuum polarization, $E_c = m_e^2 c^3/(e \hbar) \approx 1.32\times 10^{16}$ V cm$^{-1}$, generating an $e^+e^-$ plasma. The plasma self-accelerates owing to its internal pressure, loads with it some baryons from the environment, and finally reaches transparency in an ultrarelativistic regime with characteristic Lorentz factor $\Gamma \sim 100$ \cite{2001A&A...368..377B, 2021PhRvD.104f3043M}. The UPE is the first manifestation of the BH and the blackbody (BB) component from the plasma transparency at MeV energies is the signature in the spectrum that allows its identification in the GRB data. Another special signature of the UPE is its \textit{hierarchical} structure shown for the first time in GRB 190114C \cite{2021PhRvD.104f3043M}, i.e., a refined time-resolved analysis of the UPE shows that its spectrum in rebinned time intervals (up to a fraction of second) shows always a cutoff power-law + blackbody (CPL+BB) model. Numerical simulations of the QED physical process for the UPE of GRB 190114C, which extends from the rest-frame time $t_{\rm rf}=1.99$~s to $t_{\rm rf}=3.99$~s, show that the plasma transparency occurs in pulses in a nanosecond timescale, which explains the similar spectra of the UPE hierarchical structure (see \cite{2021PhRvD.104f3043M} for details). In GRB 180720B, the UPE has been identified in two time intervals \cite{Rastegarnia2022-co}. The UPE I extends from $~t_{\rm rf}=4.84$~s to $~t_{\rm rf}=6.05$~s, has isotropic energy $E_{\rm UPE I}=(6.37\pm0.48) \times 10^{52}$~erg, and its spectrum is best fitted by a CPL+BB model, index $\alpha=-1.13$, cutoff energy $E_{\rm c}=2220.57$~keV, and BB temperature $k T = 50.31$~keV in the observer's frame. In the UPE II continues the UPE phase from $~t_{\rm rf}=9.07$~s to $~t_{\rm rf}=10.89$~s, has an isotropic energy of $E_{\rm UPE II}=(1.6 \pm 0.95) \times 10^{53}$~erg, and its spectrum is best fitted by a CPL+BB model with $\alpha= -1.06$, $E_ c=1502.5$~keV, and $kT= 39.8$~keV. The UPE of GRB 180720B also shows the hierarchical structure in rebinned time intervals first observed in GRB 190114C. The electric energy that powers the plasma is induced by the gravitomagnetic interaction of the BH and the magnetic field, so the BH extractable energy powers the UPE. Each process of expansion and transparency of the plasma takes away a fraction of mass and angular momentum of the BH. The UPE ends when the induced electric field becomes undercritical. For GRB 190114C, it occurs at $t_{\rm rf}=3.99$~s \cite{2021PhRvD.104f3043M}, while for GRB 180720B, at $t_{\rm rf}=10.89$~s \cite{Rastegarnia2022-co}.

{The UPE is similar to the emission of the jet in the traditional fireball model in which the a collimated relativistic jet expands with $\Gamma \sim 10^2$--$10^3$ (see, e.g., Refs. \cite{1990ApJ...365L..55S, 1992MNRAS.258P..41R, 1993MNRAS.263..861P, 1993ApJ...415..181M, 1994ApJ...424L.131M}). One of the main differences between this model and the UPE in the BdHN scenario is the duration of this emission. The jetted fireball continues to emit while the central engine powers it, so the internal and external shocks keep interacting with the interstellar medium extending the emission from the prompt to the afterglow, including the very-high-energy emission by synchrotron self-Compton radiation (see, e.g., \cite{2002ARA26A..40..137M, 2004RvMP...76.1143P, 2019Natur.575..455M, 2019Natur.575..448Z}, and references therein). In the BdHN model, the UPE occurs only while the induced electric field is overcritical and can create the $e^+e^-$ plasma. These conditions in the BdHN last short (few seconds) and explain only the prompt emission of the GRB. There are no additional mechanisms to produce $e^+e^-$ pairs, so when the electric field becomes undercritical, the UPE shuts down. The induced undercritical field keeps extracting the BH energy powering the GeV afterglow emission by synchrotron radiation of accelerated electrons (see details in \cite{2019ApJ...886...82R, 2020EPJC...80..300R, 2021A&A...649A..75M, 2022ApJ...929...56R}). The synchrotron radiation from the expanding ejecta of the SN powered by the emission of the $\nu$NS explains the X-optical-radio afterglow (see below for further details).} Therefore, the emission of the $e^+e^-$ ultrarelativistic ($\Gamma \sim 100$) plasma is limited to the UPE and does not contribute to the GRB afterglow emission.

{
\textit{\textbf{The $\nu$NS-rise}}. The accretion of ejecta onto the $\nu$NS and the NS companion transfer mass and angular momentum to them. One-dimensional simulations of the above process has been presented in \cite{2014ApJ...793L..36F, 2015PhRvL.115w1102F}, two-dimensional in \cite{2015ApJ...812..100B}, and three-dimensional in \cite{2016ApJ...833..107B, 2019ApJ...871...14B}. Since the magnetic field of the $\nu$NS is expected to be larger than the one of the older NS companion, we expect the $\nu$NS to dominate the observed energy release in this phase. In GRB 190114C, the $\nu$NS-rise emission extends from $t_{\rm rf}=0.79$ s to $t_{\rm rf}=1.18$ s \cite{2021PhRvD.104f3043M, 2021MNRAS.504.5301R}. In GRB 180720B, it extends from $~t_{\rm rf}=6,05$~s to $~t_{\rm rf}=9.07$~s, has an isotropic energy of $E_{\nu \rm NS}=(1.13\pm0.04) \times 10^{53}$~erg, and its spectrum is best fitted by a CPL model ($\alpha=-0.98$, and  $E_{\rm c}=737$~keV, in the observer's frame). The energy released from the $\nu$NS-rise becomes dominant over the UPE for about $3$ s, which explains the observed apparent split UPEs I and II discussed above. After that time, the $\nu$NS-rise emission fades and the UPE becomes again observable. Recent numerical simulations of the early evolution of BdHN I (Becerra et al., submitted; see also \cite{2019ApJ...871...14B}) show that the NS companion can reach the critical mass for BH formation before the second peak of fallback accretion experienced by the $\nu$NS. This phenomenon makes indeed possible for the $\nu$NS-rise emission to superpose to the UPE in some cases.
}

{
\textit{\textbf{The Cavity}}. The massive accretion process onto the NS companion and the BH formation reduce the matter density around the newborn BH \cite{2019ApJ...871...14B}. Numerical simulations show that the expanding $e^+e^-$ plasma causes a further decrease of the density from $10^{-7}$ g cm$^{-3}$ to a value as low as $10^{-14}$ g cm$^{-3}$, and its interaction with the cavity walls generates emission characterized by a spectrum similar to a Comptonized blackbody with a peak energy of a few hundreds of keV \cite{2019ApJ...883..191R}. For GRB 190114C, the emission from the cavity extends from $t_{\rm rf} = 11$ s to $20$ s \cite{2021PhRvD.104f3043M}. For GRB 180720B, it occurs from $t_{\rm rf}=16.94$~s to $~t_{\rm rf}=19.96$~s, with an isotropic energy of $E_{\rm CV}^{\rm MeV}=(4.32 \pm 0.19) \times 10^{52}$~erg, characterized by a CPL spectrum ($\alpha=-1.16$, $E_{\rm c} = 607.96$~keV).
}

{
\textit{\textbf{Soft X-ray flares (SXFs) and hard X-ray flares (HXFs)}}. In the regions of high matter density surrounding the newborn BH site, the expanding $e^+e^-$ plasma engulfs high amounts of baryons leading to transparencies occurring at distances $\sim 10^{12}$ cm with Lorentz factors $\lesssim 5$, observable as SXFs and/or HXFs (see \cite{2018ApJ...869..151R} for numerical simulations and specific examples). The HXF of GRB 180720B occurs from $t_{\rm rf}= 28.95$~s to $t_{\rm rf}= 34.98$~s, with $L_{\rm HXF,iso}=(7.8 \pm 0.07) \times 10^{51}$~erg~s$^{-1}$, and its spectrum is best fitted by a CPL model with $E_c=(5.5_{-0.7}^{+0.8}) \times 10^2$~keV, $\alpha = -1.198 \pm 0.031$. The SXF occurs from $t_{\rm rf}= 55$~s to $t_{\rm rf}= 75$~s, with $L_{\rm SXF,iso}=1.45\times 10^{50}$~erg, and its spectrum is best fitted by a PL+BB model with $\alpha = -1.79 \pm 0.23$, and $k T=0.99 \pm 0.13$~keV.
}

\begin{figure*}
\centering
\includegraphics[width=0.49\hsize,clip]{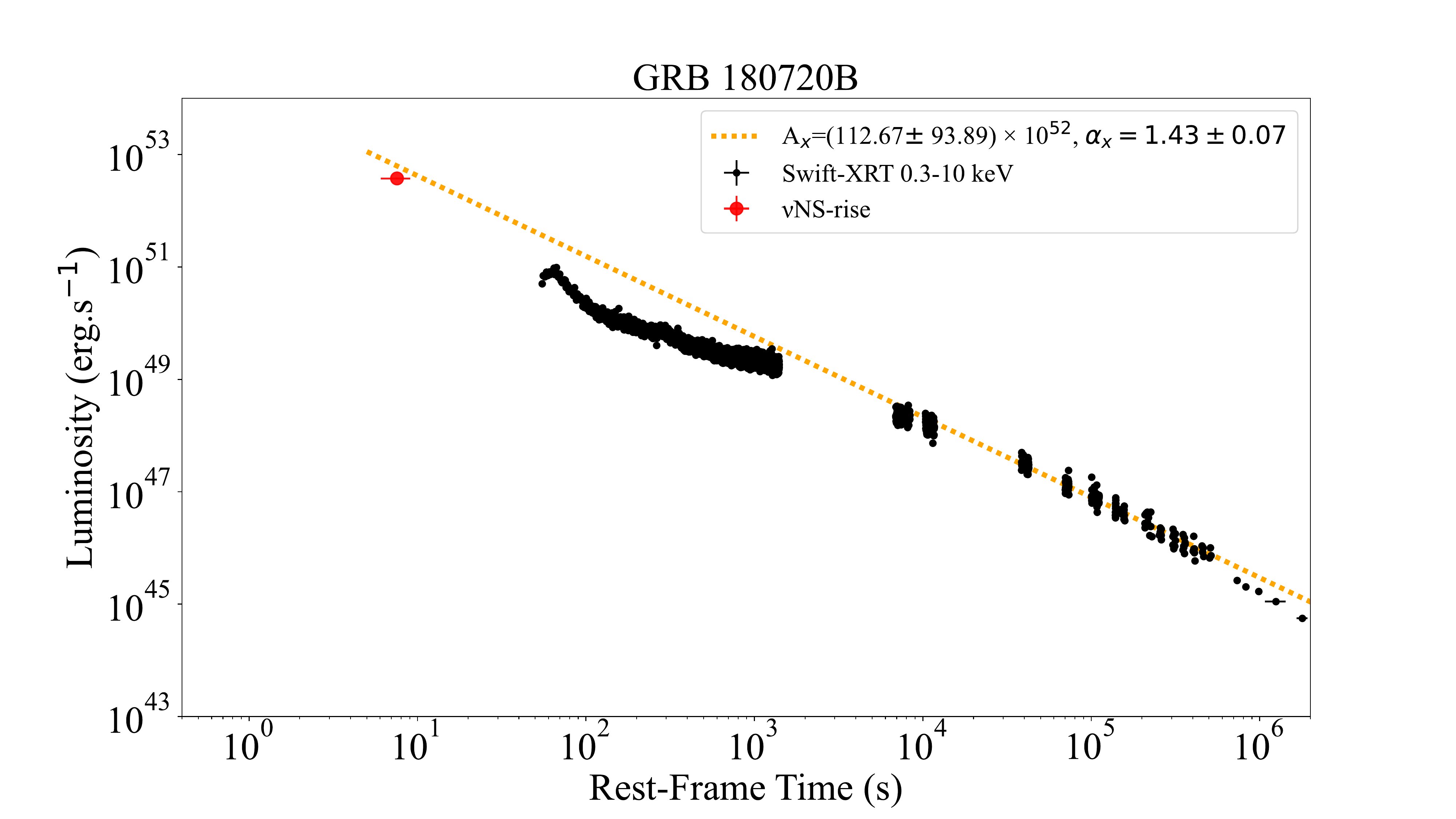}
\includegraphics[width=0.49\hsize,clip]{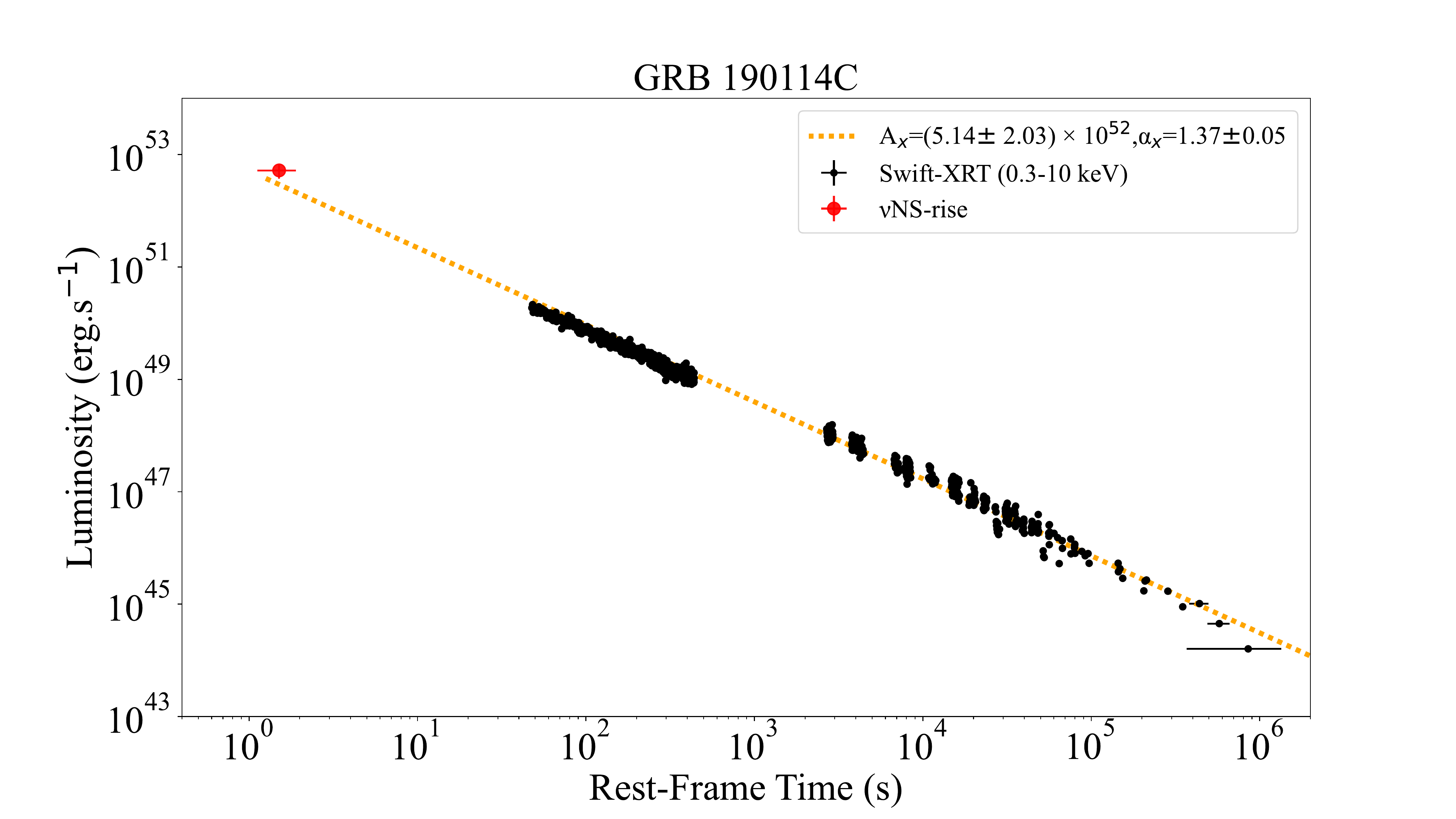}
\caption{Luminosity of the $\nu$NS-rise (red point) and the X-ray afterglow (black points) of GRB 180720B (left panel) and GRB 190114C (right panel) observed by \textit{Swift}-XRT, measured in the cosmological rest-frame. The dashed yellow line is the power-law fit given by Eq. (\ref{eq:lumx}).}
\label{fig:lumxrt}
\end{figure*}

{
\textit{\textbf{The X-ray afterglow.}}} In \cite{2021MNRAS.504.5301R}, $380$ {long GRBs have been identified as BdHN I. It has been there shown that their} X-ray afterglow, observed by the Neil Gehrels Swift satellite \cite{2016ApJ...832..136R, 2016ApJ...833..159P, 2018ApJ...852...53R}, with a luminosity in the cosmological source rest-frame that decreases with time as a power-law \cite{2021MNRAS.504.5301R}, i.e.,
\begin{equation}\label{eq:lumx}
        L_X=A_X t^{-\alpha_X},
\end{equation}
where $A_x$ and $\alpha_X$ depend on the source. In this article, we confine our attention to the BdHN I prototypes GRB 180720B, with $A_X= (2.5\pm 0.4)\times 10^{53}$~erg~s$^{-1}$ and $\alpha_X=1.44\pm 0.01$, and GRB 190114C, with $A_X= (5.14\pm 2.03)\times 10^{52}$~erg~s$^{-1}$ and $\alpha_X=1.37 \pm 0.05$. Figure \ref{fig:lumxrt} shows the $\nu$NS-rise and the X-ray afterglow of GRB 180720B and GRB 190114C, in the cosmological rest-frame of the sources. In GRB 180720B, the $\nu$NS-rise is observed at $6.05$--$9.06$~s, and in GRB 190114C at $1.12$--$1.68$ s from the Fermi-GBM trigger. The afterglow in the optical and radio energy bands also shows a similar power-law but much less luminous, so the X-ray luminosity is an excellent proxy of the total {(bolometric)} afterglow luminosity{, i.e., we assume $L_\infty \approx L_X$}.

{In the BdHN model, the GRB afterglow is explained by the electron synchrotron radiation produced in the SN ejecta while it expands through the $\nu$NS magnetic field lines, and the $\nu$NS pulsar emission that becomes observable in the late-time afterglow (see, e.g., \cite{2018ApJ...869..101R, 2020ApJ...893..148R}). Numerical simulations of the hypercritical process onto the $\nu$NS \cite{2019ApJ...871...14B} show that it gains sufficient energy (and angular momentum) during the early fallback accretion to power the energy of the observed afterglow. The fit of the X-ray afterglow data with the above synchrotron radiation model shows that typically the magnetic field at $\sim 10^{12}$~cm is $B \sim 10^5$~G and decreases linearly with the radial distance. This behavior is indeed expected from the toroidal component of the $\nu$NS magnetic field at large distances from the light cylinder \cite{2018ApJ...869..101R, 2019ApJ...874...39W, 2020ApJ...893..148R}. Summarizing, the synchrotron emission occurs in the optically thin region of the SN ejecta that expands at mildly-relativistic velocity, $v\approx 0.1c$, in the $\nu$NS magnetic field, at distances above $10^{12}$~cm. We refer the reader to \cite{2020ApJ...893..148R} for the application of the above afterglow model to GRBs 130427A, 160509A, 160625B, 180728A, and 190114C.}

\section{$\nu$NS structure evolution}\label{sec:III}

{In Figure \ref{fig:lumxrt}, we show the backward extrapolation to early times of the power-law luminosity of the X-ray afterglow. We notice that it joins the $\nu$NS-rise emission. We interpret this coincidence as an observational verification of the BdHN picture that the $\nu$NS energy powers the $\nu$NS-rise and the afterglow emissions. The above is our central working hypothesis in this article. Therefore, we assume the $\nu$NS-rise is the first release of the $\nu$NS energy gained during the fallback accretion process and continues to release it at the pace given by the power-law luminosity inferred from the X-ray afterglow.}

We are not here interested in the precise modeling of the emission mechanisms but in estimating the $\nu$NS parameters and their evolution, consistent with the required energetics at every time. In this way, we avoid including ad-hoc models for the radiation mechanism and the removal of energy and angular momentum. {For instance, the traditional model of magnetic-dipole radiation might not be sufficient for an accurate} description of the rotational energy loss of pulsars. {The measurements of the pulsar braking index deviate from} the expected value ($n=3$) of magnetic dipole radiation (see, e.g., \cite{2015MNRAS.446..857L}). {Deviations from the pure dipole braking in the very-early life of pulsars can be due to the occurrence of glitches (see \cite{2015MNRAS.446..857L, 2021MNRAS.508.3251L, 2022MNRAS.511.3304M} and references therein) which could also release high-energy emission observable in GRBs (see, e.g., \cite{2021arXiv210309158M}). In addition,} the explanation of the late-time afterglow of GRBs demands at least the presence of a substantial quadrupole component (see, e.g., \cite{2018ApJ...869..101R, 2020ApJ...893..148R}).

{
Having established all the above, we obtain the evolution of the $\nu$NS from the energy conservation equation
\begin{equation}
\label{eq:LumdEdt}
\dot{E} = -L_\infty \approx - L_X,
\end{equation}
where $L_X$ is given by Eq. \eqref{eq:lumx} and we assume it as valid from the $\nu$NS-rise time on. In agreement with the afterglow description in the BdHN model, we do not apply a beaming correction to the required $\nu$NS energetics, and do not include gravitational-wave radiation losses since the $\nu$NS is axially symmetric in this phase. {In Eq. (\ref{eq:LumdEdt}), we are assuming that the transient proto-NS regime in which the $\nu$NS energy loss is dominated by neutrino emission is over. Therefore, we consider the energy loss is dominated by photons and that the $\nu$NS is cold, so its energy is dominated by the kinetic rotational energy and the gravitational energy [see Eq. (\ref{eq:Etot}) below].}
}

We turn now to evaluate the $\nu$NS parameters and their evolution during the $\nu$NS-rise and the afterglow emissions. For this task, we model the $\nu$NS as a stable Maclaurin spheroid, i.e., a self-gravitating, oblate, homogeneous (i.e., uniform density), rigidly rotating Newtonian configuration of equilibrium. {We refer the reader to Ref. \cite{1969efe..book.....C} for details on these incompressible configurations, and to Ref. \cite{1993ApJS...88..205L} for the generalization to compressible polytropes.} From the solution of the gravitational Poisson equation, it turns out that given a density $\rho$, all the properties of the spheroid are function of the eccentricity, $e^2 \equiv (a^2 - b^2)/a^2$, where $a$ and $b$ are, respectively, the semi-major (equatorial) and semi-minor (polar) axis. The angular velocity is given by (e.g., \cite{1990ApJ...359..444F})
\begin{align}\label{eq:ang}
\Omega^2&=2 \pi G \rho g(e),\\
g(e)&=\frac{\left(3-2 e^2\right) (1-e^2)^{1/2}\arcsin (e)}{e^3}-\frac{3 \left(1-e^2\right)}{e^2}.
\end{align}
The angular momentum, $J$, and moment of inertia, $I$, are given by
\begin{equation}
\label{eq:J}
J=I\Omega,\quad I=I_0 (1-e^2)^{-1/3}, \quad I_0=\frac{2}{5}M a_0^2,
\end{equation} 
where the mass and equatorial radius are
\begin{equation}
\label{eq:M}
M= \frac{4\pi}{3} \rho a^3 (1-e^2)^{1/2},\quad a=a_0 (1-e^2)^{-1/6},
\end{equation}  
being $a_0$ the radius of the homogeneous, non-rotating (i.e. spherical) star of mass $M$, density $\rho$, and with the same volume of the spheroid, so it {fulfills the equation
\begin{equation}\label{eq:rho}
    \rho = \frac{3 M}{4 \pi a_0^3}.
\end{equation}
}
The total energy is the sum of the kinetic rotational ($T$) and gravitational ($W$) energy
\begin{equation}\label{eq:Etot}
E=T+W,\quad W= -\frac{3}{5} \frac{G M^2}{a} \frac{\arcsin(e)}{e},\quad T= \frac{1}{2} I \Omega^2,
\end{equation}  

Following the BdHN scenario, the $\nu$NS must cover the energy released in the $\nu$NS-rise and the subsequent afterglow emission. This is confirmed by the backward extrapolation of the X-ray afterglow power-law emission to the time of the $\nu$NS-rise, which shows the connection between the two emissions (see Fig. \ref{fig:lumxrt}).

By integrating analytically Eq. (\ref{eq:LumdEdt}), and equating it to Eq. (\ref{eq:Etot}), we obtain the following algebraic, nonlinear implicit equation whose solution gives the eccentricity as a function of time
\begin{align}
\label{eq:et}
&\pi G I_0 \rho\,{\cal F}(e) =\frac{A_X}{1-\alpha_x }t^{1-\alpha_x},\\
&{\cal F} \equiv -2 + \frac{3(1-e^2)^{2/3}}{e^2}+\frac{(4 e^2-3) (1-e^2)^{1/6}}{e^3}\arcsin(e),
\end{align}
where we have used the asymptotic condition $e(\infty) = 0$.

{Therefore, Eq. \eqref{eq:LumdEdt} and the equilibrium properties of the Maclaurin spheroid allows us to estimate the evolution of all the relevant physical properties of the $\nu$NS. The above framework tells us that given values of $M$ (or alternatively $\rho$) and $a_0$, all stellar parameters (energy, angular momentum, moment of inertia, angular velocity) depend only on the eccentricity, $e(t)$. Summarizing, the solution $e(t)$ is obtained from Eq. \eqref{eq:LumdEdt} which leads to the implicit algebraic equation \eqref{eq:et}. With the knowledge of $e(t)$, the evolution of the rotational energy and the gravitational energy are obtained from Eqs. \eqref{eq:Etot}, likewise the evolution of the radius from Eq. \eqref{eq:M}, the angular momentum and moment of inertia from Eqs. \eqref{eq:J}, and the angular velocity from Eq. \ref{eq:ang}.}

\begin{figure*}[htbp]
\centering
\begin{tabular}{cc}
\includegraphics[width=0.49\hsize,clip]{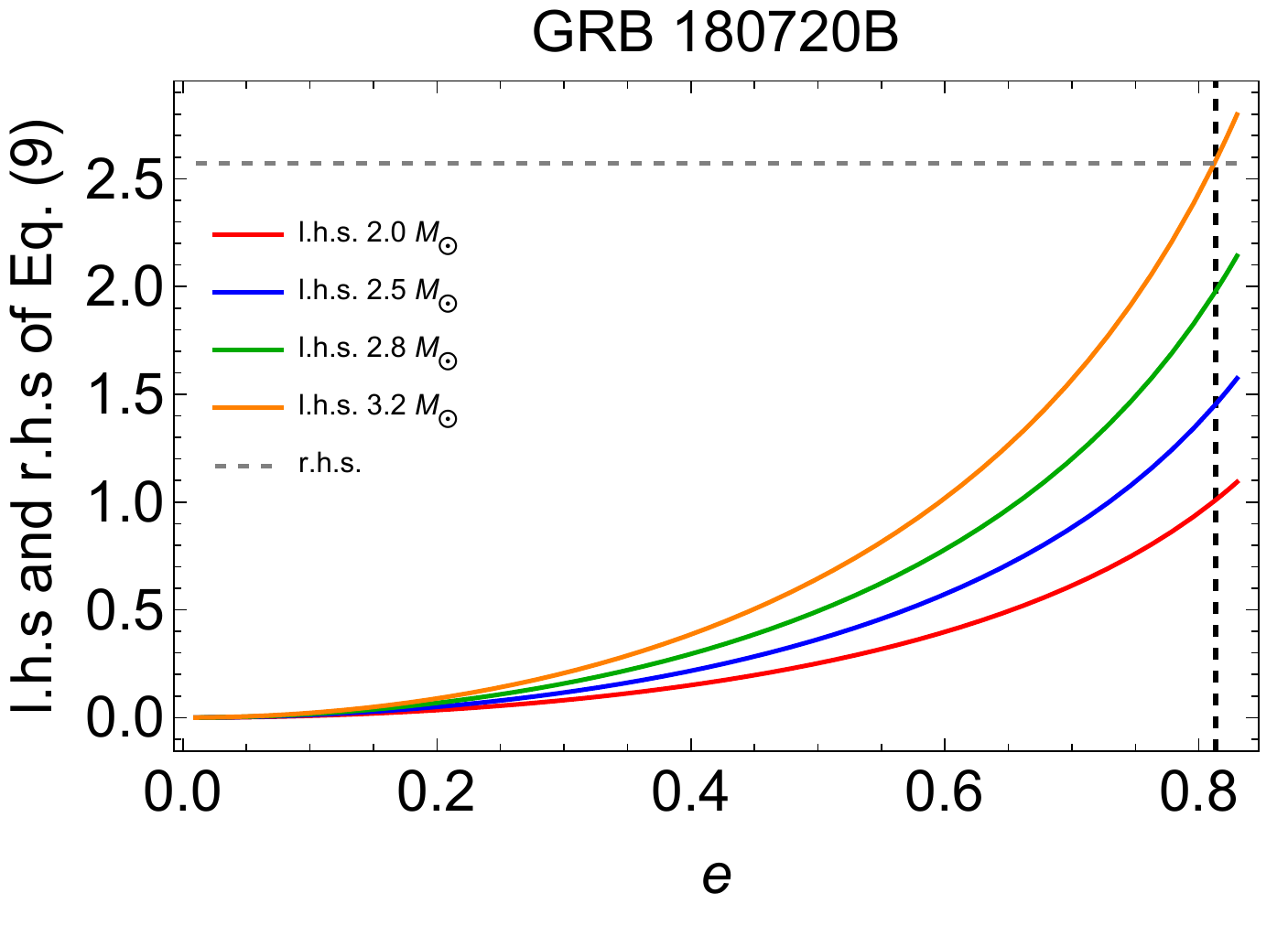} & \includegraphics[width=0.49\hsize,clip]{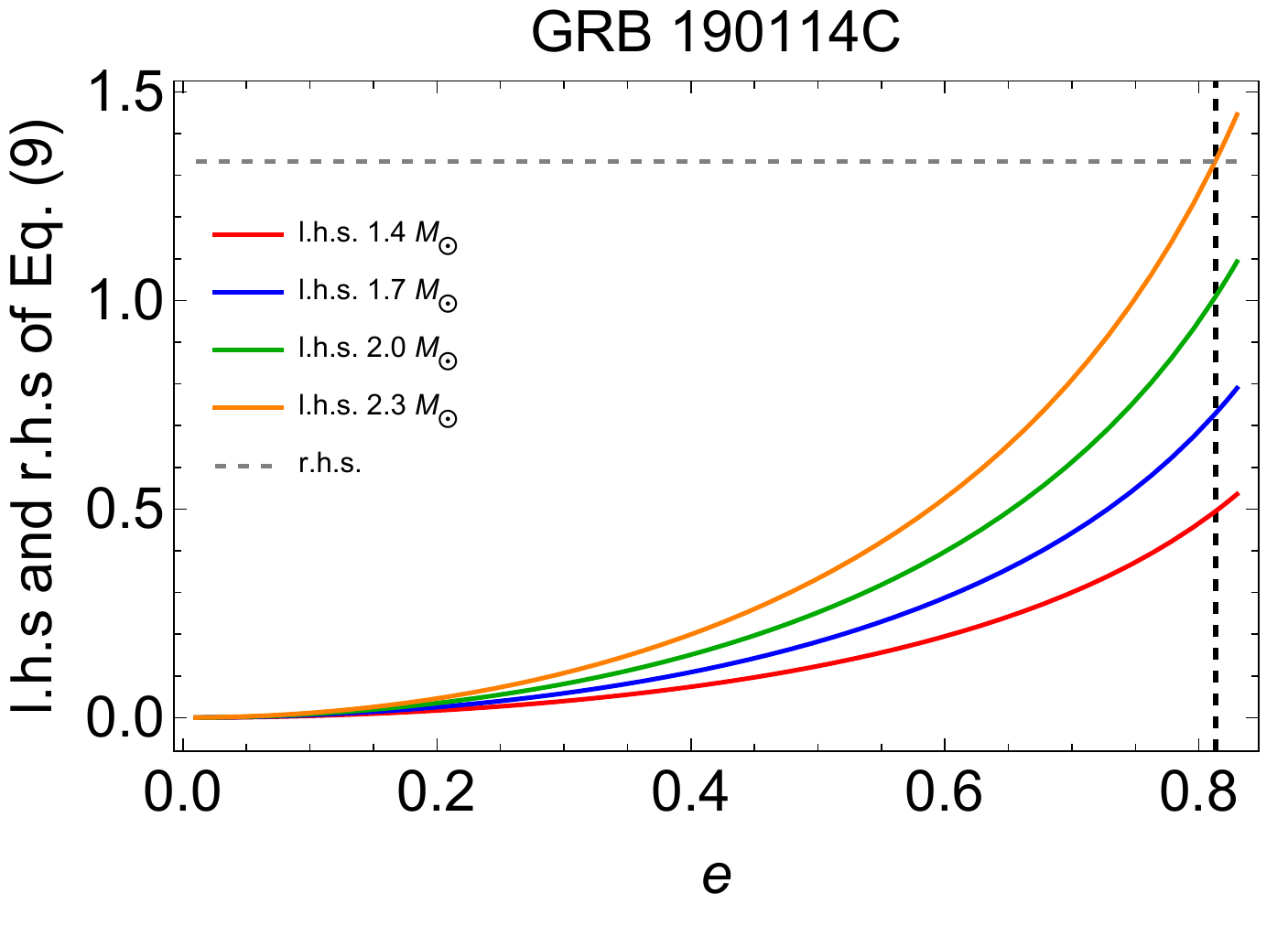}
\end{tabular}
\caption{{Left-hand side (l.h.s., colored curves) and right-hand side (r.h.s, dashed gray horizontal line) of Eq. \eqref{eq:et} at the $\nu$NS-rise time, for GRB 180720B (left panel) and GRB 190114C (right panel). The dashed black vertical line marks the maximum eccentricity of stable Maclaurin spheroids, $e_{\rm max} \approx 0.813$. The units of the vertical axis are of $10^{53}$ erg s $^{-1}$.}}
\label{fig:solmass}
\end{figure*}

As for the initial conditions, we must specify a value {of $M$ (or alternatively $\rho$)}, $a_0$, and at the initial time of evolution, $t_0$, which is the time of the occurrence of the $\nu$NS-rise, a value for the eccentricity, $e(t_0)$. We here adopt $a_0 = 10^6$ cm, and seek for the {mass} and initial eccentricity of the spheroid that allow to explain the $\nu$NS-rise and afterglow emission, which are specified by the values of $A_X$ and $\alpha_X$.

\begin{table}[htbp]
    \centering
    {
    \begin{tabular}{c|c|c}
     & GRB 180720B & GRB 190114C \\
     \hline
    $A_X$ & $(2.5\pm 0.4)\times 10^{53}$ & $(5.14\pm 2.03) \times 10^{52}$\\
    $\alpha_X$ & $1.44\pm 0.01$ & $1.37 \pm 0.05$\\
    $t_0 (s)$ & $6.05$ & $1.12$ \\
    $e_0$ & $0.813$ & $0.813$ \\ 
    $a_0$ ($10^6$ cm) & $1.0$ & $1.0$ \\
    {$a$ ($10^6$ cm)} & {$1.2$} & {$1.2$} \\
    $M (M_\odot)$ & $3.19$ & $2.3$ \\
    $\rho$ ($10^{15}$ g cm$^{-3})$ &  $1.52$ & $1.09$ \\ 
    $P_0$ (ms) & $0.58$ & $0.68$ \\
      \hline
    \end{tabular}
    }
    \caption{Properties of the $\nu$NS modeled as a Maclaurin spheroid that powers the $\nu$NS-rise and the X-ray afterglow in GRB 180720B and GRB 190114C. {The radius $a_0$ is assumed to be $10^6$ cm, and we obtain the mass $M$ seeking for the solution of Eq. \eqref{eq:et} at the initial time $t_0$ ($\nu$NS-rise time) as shown in Fig. \ref{fig:solmass}. The value reported here is the minimum mass, which corresponds to the solution for the maximum eccentricity of stable Maclaurin configurations, $e_{\rm max} \approx 0.813$. The density is given by Eq. \eqref{eq:rho}. The corresponding initial rotation period of the configuration is obtained from the initial rotation angular velocity, $\Omega_0 = 2\pi/P_0$, where $\Omega_0$ is calculated by plunging $\rho$ and $e_0$ into Eq. \eqref{eq:ang}}.}
    \label{tab:properties}
\end{table}

We have found that the high luminosity and energy released at the $\nu$NS-rise requires the $\nu$NS to have the fast spin and the maximum (or very close to it) eccentricity allowed by the axially symmetric Maclaurin spheroid, namely the values of the bifurcation point to the sequence of Jacobi ellipsoids (triaxial configurations), i.e., $e(t_0) = 0.813$ \cite{1969efe..book.....C}. Table \ref{tab:properties} summarizes the initial conditions of the Maclaurin spheroid modeling the $\nu$NS in GRB 180720B and GRB 190114C. {The inferred rotation periods correspond to frequencies of $1.72$ kHz for GRB 180720B and $1.47$ kHz for GRB 190114C. These high rotation rates are indeed close to the maximum allowed values of uniformly rotating NSs (see, e.g., \cite{2015PhRvD..92b3007C}), which is consistent with the Maclaurin spheroid be at the bifurcation with the Jacobi sequence. The fallback accretion process in the first minutes of the $\nu$NS life can transfer sufficient mass and angular momentum to bring it to these critical values (see \cite{2019ApJ...871...14B} and \cite{2022arXiv220803069B} for recent numerical simulations of this process in BdHNe).}

\begin{figure*}[htbp]
\centering
\begin{tabular}{cc}
\includegraphics[width=0.44\hsize,clip]{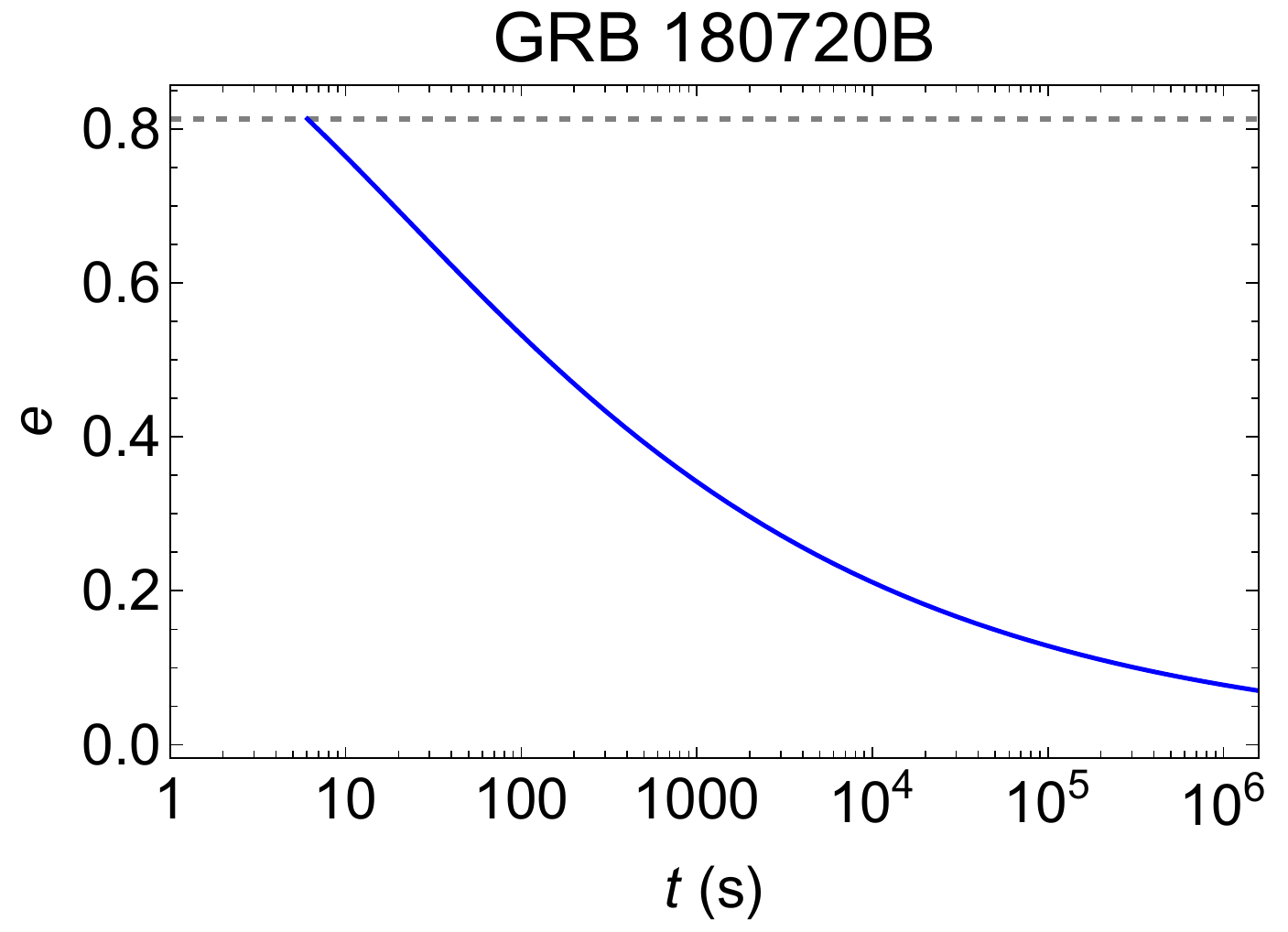} & 
\includegraphics[width=0.44\hsize,clip]{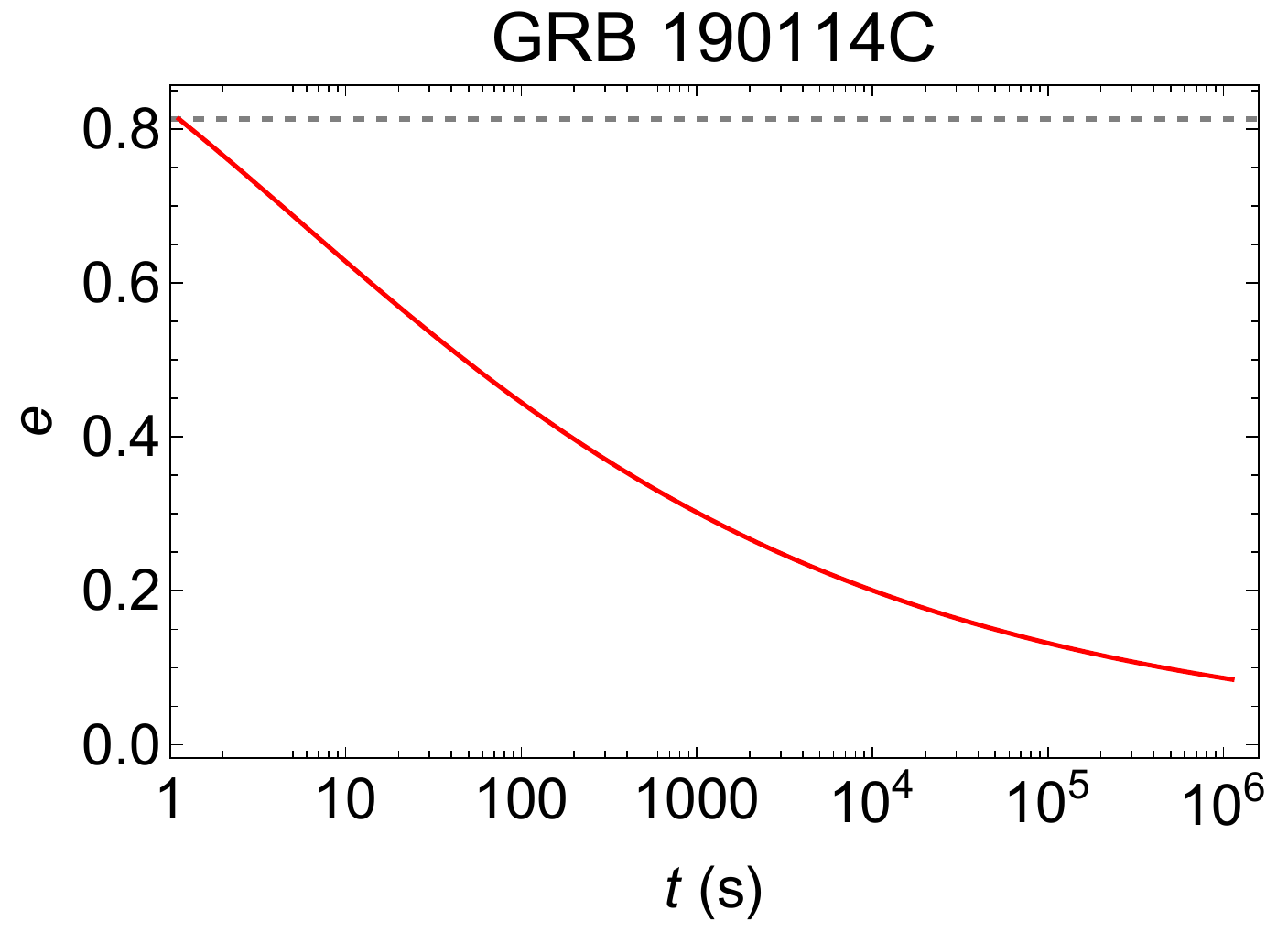}\\
\includegraphics[width=0.44\hsize,clip]{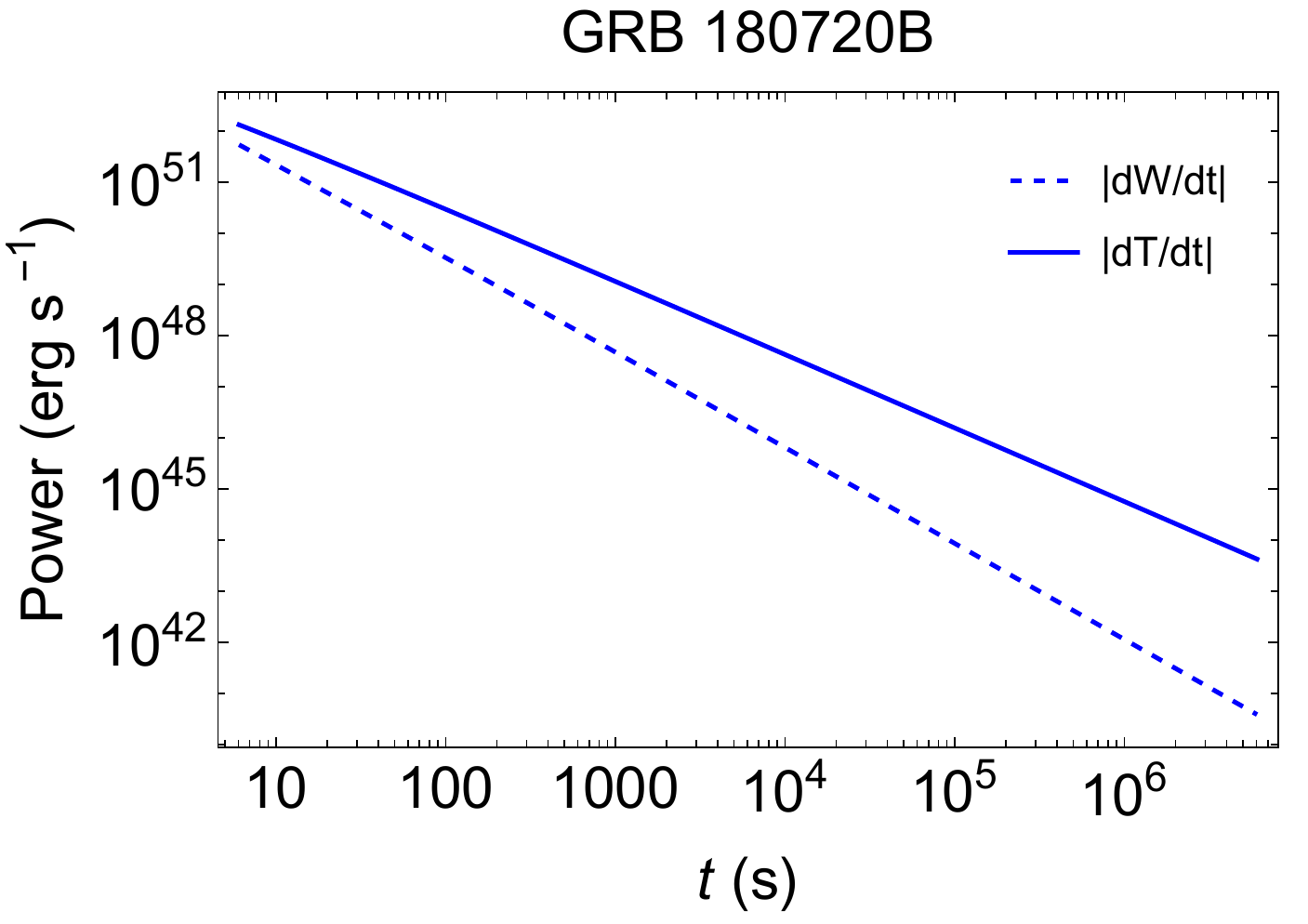} &
\includegraphics[width=0.44\hsize,clip]{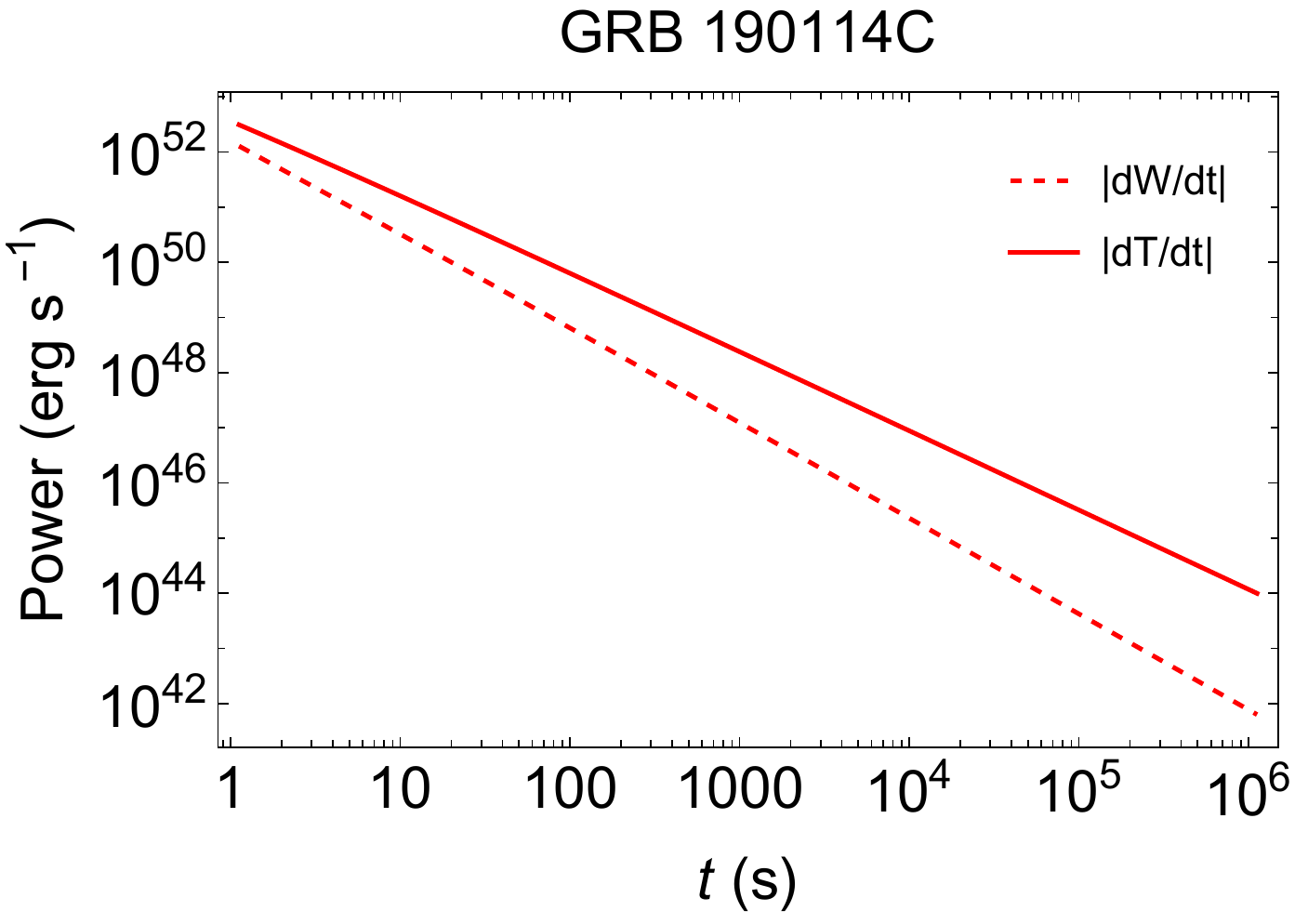}\\
\end{tabular}
\caption{Evolution of the eccentricity (upper row) and the gravitational and rotational power released (lower row) by the $\nu$NS modelled as a Maclaurin spheroid, in the cases of GRB 180720B (blue curves) and GRB 190114C (red curves).
}
\label{fig:evst}
\end{figure*}

Figure \ref{fig:evst} shows the evolution of the eccentricity (upper row) and the contribution of the rotational and gravitational power (lower row) to the total power released during the evolution. {The rotational power dominates over the gravitational power during most of the evolution, although the latter contributes significantly at early times. For instance, $\dot{W}/\dot{E}\gtrsim 0.1$ at eccentricities $e\gtrsim 0.5$.}

\section{Concluding remarks}\label{sec:IV}

We have calculated the evolution of the $\nu$NS in BdHN I assuming that it powers the $\nu$NS-rise and the afterglow emission. By modelling the $\nu$NS as a Maclaurin spheroid, we have shown that its parameters (rotation period and eccentricity) have to be very close (or equal to) to the ones of the transition point to the Jacobi sequence of ellipsoids (see Table \ref{tab:properties}).

At the bifurcation point with the Jacobi ellipsoids sequence, the ratio of the rotational to gravitational energy of Maclaurin spheroids is $T/|W| \approx 0.14$, where the configuration becomes secularly unstable and is evolution driven by gravitational radiation \cite{1969efe..book.....C}. Since the $\nu$NS at the $\nu$NS-rise time are at, or close to, the bifurcation point, they have this $T/W$ ratio. In addition, assuming a spherical radius of $a_0=10^6$ cm, we have found that the mass of the spheroids is $3.2 M_\odot$ and $2.3 M_\odot$, respectively, for GRB 180720B and GRB 190114C (see Table \ref{tab:properties}). Centrally condensed objects (so not of uniform density) at rigid rotation become unstable against mass shedding (Keplerian limit) at lower values of this ratio (see, e.g., \cite{2015PhRvD..92b3007C}). Instead, configurations with differential rotation can have a $T/|W|$ ratio up to the maximum value of $0.5$ set by the virial theorem. Given the mass and $T/|W|$ ratio, our result suggests that the $\nu$NS might have some differential rotation. Therefore, the $\nu$NS could evolve from the hypermassive stability region (mass $>$ maximum mass of rigidly rotating stars; supported by differential rotation) into the supramassive one (mass $>$ maximum mass of non-rotating stars; supported by rigid rotation). The above situation is similar to the evolution of a hypermassive NS formed in a NS binary merger (see, e.g., \cite{2020ApJ...902L..41R, 2022MNRAS.509.1854R}, for the stability analysis of the merged object modeled as a Riemann-S ellipsoid) and becomes an interesting topic of further investigation in a fully general relativistic framework (see, e.g., \cite{2017PhRvD..95l4057T}). 

{The parameters of the NS inferred from the present model suggest the properties of the $\nu$NS even if the exact values can be slightly different depending on factors like the nuclear equations of state, the interior rotation law (i.e., uniform or differential rotation), and the use of general relativity. Our analysis indicates that the NS powering the afterglow emission of these GRBs must be massive (likely $\sim 2 M_\odot$), fast rotating (likely $\sim 1$ kHz), and initially with high eccentricity. Even though, in the present model, the NS is stable since it belongs to the stable branch of Maclaurin spheroids. The high value of the NS mass is comparable to or higher than the critical mass of uniformly rotating NS in general relativity for some nuclear equations of state. It suggests the NS equations of state must be stiff, which is consistent with the observation of stable massive NSs above two solar masses, e.g., PSR J0952-0607, the heaviest NS measured to date with a mass of $2.35\pm 0.17 M_\odot$ \cite{2022ApJ...934L..18R}.}

Therefore, the $\nu$NS could have evolved from a triaxial body (Jacobi-like ellipsoid) into an axially symmetric body (Maclaurin spheroid) by emission of gravitational waves, as anticipated in early models of pulsars (e.g. \cite{1969ApJ...157.1395O, 1969ApJ...158L..71F, RWlincei1970}) and verified by \cite{1970ApJ...161..571C, 1974ApJ...187..609M}. The gravitational-wave emission drives the evolution of the ellipsoid to the Maclaurin sequence in relatively short time \cite{RWlincei1970, 1974ApJ...187..609M}. This emission carries out angular momentum which plays a role in bringing the $\nu$NS to the observed short-rotation period: the ellipsoid spins up while it loses angular momentum because the gravitational-wave-driven evolution occurs along the Riemann-S sequence conserving circulation \cite{1974ApJ...187..609M}. The gravitational-wave power released by the triaxial configuration with equatorial ellipticity $\epsilon$ and moment of inertia $I$ about the rotation axis is $\dot{E}_{\rm GW} = (32/5) (G/c^5) I^2 \epsilon^2 \Omega^6$ \cite{1969ApJ...158L..71F}. For instance, assuming a rotation frequency of $1$ kHz, and the moment of inertia inferred for the $\nu$NS in GRB 180720B, we obtain $\dot{E}_{\rm GW} \sim 1.5\times 10^{53} (\epsilon/0.1)^2$ erg s$^{-1}$, and the characteristic timescale $\tau_{\rm GW} \sim E/\dot{E}_{\rm GW} \lesssim 1$ s, where $E\sim 10^{53}$ erg is the gravitational energy of the triaxial configuration, and we are assuming that the ellipticity can be as large as $0.1$ at early post-birth times. {This implies a large amount of energy carried out by this \textit{burst} of gravitational waves, $\Delta E_{\rm GW} \sim \dot{E}_{\rm GW} \tau_{\rm GW}\sim 10^{53}$ erg}. The associated characteristic strain at a detector of gravitational waves is $h_c \sim 4 G/(c^4 D) I \epsilon \Omega^2 \sim 1.6 \times 10^{-23}\ (\epsilon/0.1)\ ({\rm 100\ Mpc}/D)$, where $D$ is the distance to the source (e.g. \cite{1995ApJ...442..259L}). {This signal could be detected by upgraded versions of the Nautilus cryogenic detector, which was conceived for this aim (see, e.g., \cite{2005PhRvD..71d2001A}), and working in coincidence with the} Advanced LIGO and Virgo interferometers at these frequencies (e.g., \cite{aligo2015}). In view of the above, and the possible enhancement of the strain depending upon the number of cycles of the signal in the detector \cite{1995ApJ...442..259L}, there is a chance  to calibrate gravitational-wave detectors \cite{2021PhRvD.103b2002G} observing this radiation before the GRB prompt emission for sources located at $D\lesssim 100$ Mpc. 

{In the present BdHN I scenario, the above is the specific emission of gravitational waves associated with the long GRB. The core-collapse leading to the $\nu$NS radiates negligible gravitational waves, $\Delta E_{\rm GW}\sim 10^{-7} M_\odot c^2 \sim 10^{47}$ erg (see, e.g., \cite{2002A&A...393..523D, 2011LRR....14....1F}). Since there is no relativistic jet launch in the BdHN scenario, mechanisms such as the gravitational-wave emission from an accelerating jet \cite{2021PhRvD.104j4002L} are not expected either to be at work in BdHNe \cite{2022ApJ...929...56R}.}

\begin{acknowledgments}
We thank the referees for insightful comments and suggestions that helped us to improve the presentation of the article. J.F.R. thanks financial support from the Patrimonio Aut\'onomo - Fondo Nacional de Financiamiento para la Ciencia, la Tecnolog\'ia y la Innovaci\'on Francisco Jos\'e de Caldas (MINCIENCIAS - COLOMBIA) under the grant No. 110685269447 RC-80740--465--2020, project 6955.
\end{acknowledgments}


%

\end{document}